# scientific reports

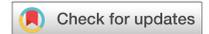

OPEN       Towards a quantum synapse for quantum sensing

L-F Pau

As a step in the architectural design of a quantum processing or sensing system with control and signaling, an attempt is made at putting in parallel functional properties of the random flows between neurons through electrical synapses, and quantum particle flows inside a quantum processing system mimicking biological processes. Based on a simplified dynamic electrical synapse model, a quantum synapse circuit design is proposed. This is extended to the case of bidirectional flows through a synapse, highlighting the possible role of quantum synapse circuits as highly parallel controlled interfaces crucial in sensing and sensor fusion systems. A short status of the quantum simulation is provided.

Whereas quantum physics[1], quantum computing[2] and quantum sensors[3,4], have grown in importance, an innovative stream of research has emerged rooted in architectural features mimicking biological elements and fluxes[5,6]. It has been motivated by three concurrent evolutions ; the first relies in exploiting new particle physics effects (such as : Moiré patterning on graphene structures[7], quantum Hall effect[8], magnetic control of quantum pathways[9], etc. ) discovered in some materials ; the second relies on the need for quantum computing to be able to address less recursive algorithmic computations and in particular embedded uses with sensing and actuation[10]; the third relies on the well-known fact that biological processes (such as: vision analysis, allostatic load, cortical reflections ) demonstrate amazing sensing and adaptation capabilities reminding of superposition and entanglement[11], especially so, as such biological processes are triggered by randomness, although of a biochemical nature very different from quantum particles. This gives the motivation justified in this paper to attempt to mimic some of these biological processes by suitable quantum circuits, even if the time scales are completely different.

More recently, it has been shown how signaling and control can be added in quantum computing architectures inspired by biology, as required in time critical embedded applications, while preserving decidability[12].

Whereas such prior results focused in more detail on the particle physics effects exploited in early realizations[5], as well as on biological experiments supporting the developed architecture[13](such as combined light and sound excitations on living creatures (Patent [A]), the present paper illustrates how biological processes around a synapse can lead to the specification of quantum circuits with high relevance in some quantum sensing and sensor integration applications. To this end, the choice of electrical synapses as a base is driven by their capability to interface and conduct selective fusion between specialized neurons or specialized quantum sensors. It is to be recalled, that neurons offer native spatiotemporal integration[14], correlation[15–17], as well as gating, interference filtering, splitting and other functionalities required in sensor signal processing; for an overview in biology see[18].

This research rests on extensive research in biology and on quantum processing architectures inspired by biology. It is necessary to point out upfront that artificial neural networks and their associated theoretical models, are out of scope and of little relevance here, although some attempts have been made at conducting the training of artificial neural networks on quantum computers[19]. Indeed, the approach is not to exploit mathematical models for an artificial neuron or synapse, but to embed the biological signal behavior directly into quantum gates.

## Research question

The goal of this paper is to propose a quantum circuit design of an electrical synapse, rooted in biological interaction processes and signals exchanged with neurons, designated under the term "quantum synapse". This design is conducted in stages, from a base model used in quantitative neurology, to a quantum model with two-way flows across the synapse, adding in later the signaling & control functionality. This offers a building block towards the capability to control a network of synapses[20]. One potential application to sensor fusion is presented.

CBS (Copenhagen, Denmark), Erasmus University (Rotterdam, Netherlands), and Upgötva AB (Stockholm, Sweden), Rotterdam, Netherlands. email: lpau@nypost.dk

The intent is not to discuss here in detail a physical quantum synapse implementation with the required semiconductors, physics, and biology, but to offer a concrete vision of their interdependencies, and relevance, towards a quantum sensing realization. It should be stressed that such a physical implementation is a special case of the architecture already provided in detail in[5].

The title of this submission starting with "Towards..." stresses that this research is exploratory although meaningful results have been obtained at theoretical, experimental and simulation stages.

Section Electrical synapse connections between neurons reviews synapse connections between neurons, and points at some properties transferable at the functional level to quantum processing. For subsequent quantum circuit specification, section Base model of an electrical synapse (gap junction) provides the dynamic model of a synapse with two connected neurons. This allows in Section Quantum computing relevance for modelling an electrical synapse to summarize the relevance of electrical synapse models for quantum processing and sensing, resting on earlier experimental results[13], leading in Section Quantum synapse to the corresponding qubit operations and a base quantum synapse circuit specification. Prior results on signaling and control are applied to this quantum synapse specification in Section Control and signaling in a quantum synapse, which also addresses a specification of a quantum synapse with bilateral flows. In Section Quantum sensing evaluation application, the significance thereof in sensor fusion is discussed, as quantum processing power associated to quantum synapses may remove some fusion bottlenecks, raising issues of scalability in Section Discussion on scalability. Quantum simulations of the quantum synapse have been carried out, and the approach and tools are summarized in Section Quantum synapse simulation before a companion paper will appear. After formalizing the comparison of randomness in neurons and quantum synapse designs, and addressing other open research issues in Section Limitations, results and open research issues, a Conclusion answers the research question and further reflects on scalability.

## Electrical synapse connections between neurons

The purpose of this Section Electrical synapse connections between neurons is to analyze the electrical flows to, from and inside synapses, in order to establish the corresponding quantum flows inside a quantum-based synapse model. The randomness of the activations, and resulting biological signals inside a synapse, result from biochemical ion interactions summarized below.

Neurons communicate between themselves via synapses (see Fig. 1), playing the role of junctions, in that a given neuron, once activated by random or nerve control signals, transmits a stimulating or inhibiting message (called neurotransmitter, to a target neuron (also called postsynaptic neuron) within a binding period called synaptic delay (0,5 – 1 milliseconds)[21]. Biology identifies three different classes of synapses: spiking chemical, non-spiking chemical, and electrical synapses; the synaptic transmission can be either electrical, or chemical, or both. Each category may have an unlimited number of more specific types, all differing through the quantitative values of their parameters.

In the electrical synapses, ions flow between the cells. Electrical currents flow from one neuron to another via a synapse each time (i) there is a membrane potential difference between the two neurons, and (ii) the electrical junctional synapse conductance is greater than zero. The emitting neuron (also called pre-synaptic neuron) maintains a rest potential across its cellular membrane (about −60/−70 mV), but can also emit random nervous spikes, also called action potentials (when membrane potential rises to about −55 mV), whereby it carries out metabolic processes required by its survival (see Fig. 2). Figure 2 also displays one pre-synaptic spike, of which a sequence develops to trigger the randomness of the pre-synaptic voltage. The synapses are generally formed between nervous terminations (called axonal terminations) of the presynaptic neuron and the dendritic cells of the postsynaptic neuron. A given axonal termination may have several branches, thus allowing it to

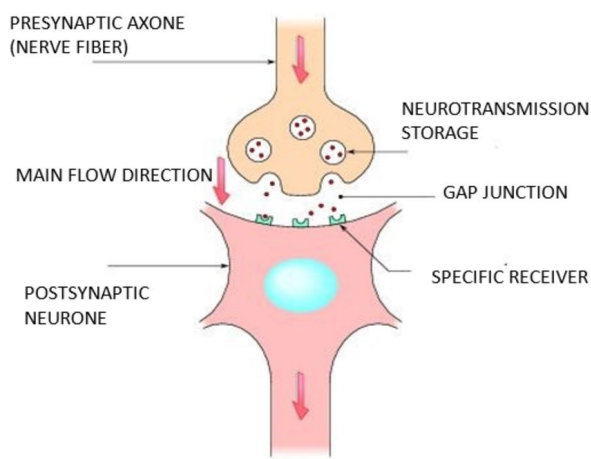

**Fig. 1.** Electrical synapse connections.

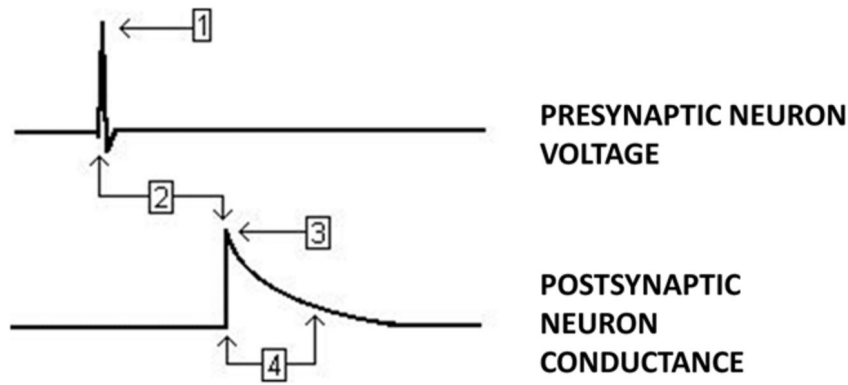

1. Pre-synaptic spike.
2. Delay between pre-synaptic spike and post-synaptic response.
3. Initial amplitude of post-synaptic conductance increase.
4. Decay rate (exponential time constant) of post-synaptic conductance increase

**Fig. 2**. Synaptic spikes (Equations 2,5).

establish connections to synapses having different postsynaptic neurons. In this way a given neuron may receive thousands of synaptic influxes originating in different presynaptic neurons through different synapses.

By the combination of random chemical postsynaptic excitation potentials which depolarize the cell, and postsynaptic inhibition potentials which lower the membrane potential, a postsynaptic chemical neuron may add or integrate all excitation flows, to decide to activate or not a binary action potential with some delay and retain some memory via plasticity[21,22]. More precisely, calcium ions Ca2 + in the presynaptic neuron access the synapse through voltage-gated channels, after the depolarization of this presynaptic neuron has happened. This influx of Ca2 + ions fills sacs with neurotransmitters to move them to the membrane facing the synaptic cleft. Such sacs fuse with the plasma membrane so that exocytosis takes place, releasing a chemical message which gets sent to the postsynaptic neurons. Dynamic cell and synapse biophysics result in differential equations to generate dynamic neurotransmission response models[23–26].

The electrical synapses, contrary to chemical synapses[27,28], realize a direct physical connection between a given presynaptic neuron and all the postsynaptic neurons, in the form of communication junctions which allow ions randomly to circulate fast (but not instantly)[29,30]. The electrical synapse has two close membranes with protein transmission channels (called connexins) allowing for direct passage of current. In those synapses which are both electrical and chemical, the electrical response occurs before the chemical response. It should be noted that the communication junctions may carry two-way flows, later studied in Section Control and signaling in a quantum synapse, so that e.g., the depolarization of a postsynaptic neuron may induce the depolarization of a presynaptic neuron, and more generally, allowing a synchronization between groups of neurons despite the randomness. As mentioned, there are many types of electrical synapses, but also parametric structures that may control the synchronization of the rhythmic activity and spike release in neuronal networks[31,32]. In addition, electrical synapses may function as low-pass filters and transfer spikes after their hyperpolarization. However, the electrical synapses cannot, like chemical ones, transform an excitation signal in a neuron into an inhibition signal in another neuron.

It results from the above by correspondence that in an electrical synapse model:

a. The realization of a quantum electrical synapse model should allow for networked quantum flow distribution, but at a vastly enhanced speed and with many more flows than with current technologies;
b. Quantum electrical synapses cannot realize a binary function; this is no problem as this functionality is devoted to quantum gates for which realizations are described in[5];
c. If some synapses are devoted to a signaling or control functionality as described in[12,] they implement in effect a quantum flow routing functionality in response to presynaptic commands (or their equivalent in the chosen quantum architecture);
d. By their bursty nature having some properties analog to spike flows (see Section Quantum computing relevance for modelling an electrical synapse), quantum particles in turn may help study random neuron-synapse-neuron interactions, as the retransmission of bursty sequences is essential in cerebellar coding[33].

## Base model of an electrical synapse (gap junction)
### Initial simple model
To model a synapse usually involves a mathematical differential equation description of the transformation of a presynaptic membrane potential difference into a postsynaptic response, for example as an ionic current[34]. The presynaptic potential difference serves as the random input impulse/spike sequence (Fig. 2, Eq. (2), Eq. (6)), and generates as output a dynamic variation in the postsynaptic membrane potential displaying randomness.

The neuron 'Leaky Integrate and Fire model' (LIF), provides a simple description of one postsynaptic neuron with one associated upstream electrical synapse (for parallelism resulting from multiple upstream presynaptic and postsynaptic neurons, see Section Electrical synapse model ):

$$Cm(dV/dt) = -gL(V(t) - Vrest) + Input(t) \qquad (1)$$

where :

- $C_m$ is the neuronal membrane capacity, approximately 1 μF/cm2[24];
- V(t) is the neurone membrane potential at time t along a given synapse link, and V(0) = membrane voltage initial value = −70,6837 mV;
- $V_{rest}$ = (−65) mV is the neurone rest state membrane potential;
- gL = membrane conductance value approximately = 0,0551 mS/cm2;
- Input(t) is the synapse current to the postsynaptic neuron, itself modelled as a random sequence of pulses (Fig. 2):

$$Input(t) = \sum_i \delta(t - ti) \qquad (2)$$

in which δ is the Dirac function, and the $t_i$'s are random impulse times, and i designates one of the impulses between times 0 and t;

• $V_{thres}$ is the threshold potential above which V(t) get's reinitialized to the value $V_{rest}$ approx. (60,5 mS/cm2, 50 mV);

so that the potential V(t) achieves an instantaneous rise when a synapse impulsion arrives, followed by an exponential decrease (Fig. 2).

The electrical synapse is in addition subject to saturation, in that it gets modified after an impulse is received from the presynapse neurone:

$$IP(t) = gs(t)(E_s - V(t)) \qquad (3)$$

in that the synaptic current is proportional to the difference between V(t) and the synaptic inversion potential $E_s$ = 0,0225 mV, and:

• gs (t) is the synapse conductance which varies with time and depends on previously received impulses from the presynaptic neuron. Each time the synapse receives such an impulse, gs(t) will grow until reaching its upper bound gsMAX, then decrease exponentially:

$$dgs(t)/dt = -gs(t)/\tau s \; subject \; to \; 0 \leqslant gs(t) \leqslant gsMAX \qquad (4)$$

where :
• τs = synaptic time constant.

### Electrical synapse model

The above initial model of Section Initial simple mode needs to be enhanced to approximate the neurological behavior of an electrical synapse in view of its quantum computing model. We still assume a biological Hogkin-Huxley formalism[35] with two reciprocally connected reticular nucleus TRN neurons (= Thalamic Reticular Nucleus), and a single electrical synapse between these two neurons. The full model[35] is here simplified to exclude chemical synapses.

The parallelism occurs twice for each given single synapse, due to parallel links to presynaptic TRN neurons (j) and separate parallel links to postsynaptic TRN neurons. To simplify the readability, additional indexes are not supplied to represent each postsynaptic link, those being parallel. This approach obviously extends easily to sets of synapses, and sets of of TRN neurons, which may feed to, and/or or receive random signals from synapses in highly combinatorial configurations.

Then in Eq. (2), Input(t) must incorporate the ion channels from all synaptic upstream links j to the postsynaptic neuron and all external presynaptic ionic currents:

$$Cm(dV/dt) = -gL\sum_j [(IPj(t) - Vrest) + Inputj(t)] \qquad (5)$$

$$Input_j(t) = \sum_i \delta(t - t_{ij}) + PresynapticInputs_j(t) + \sum_{k \neq j} gelec(V_k(t) - V(t)) \qquad (6)$$

where :

- gelec = electrical synapse conductance = 0 to 0,025 mS/cm2, with a resulting coupling coefficient of 0,2883;
- $t_{ij}$ = synapse firing timing from upstream presynaptic neuron i on link j.

### Quantum computing relevance for modelling an electrical synapse

As an introduction to this Section, it should again be pointed out that in this research, there is no direct relevance of simple artificial/computational neuron models as used for example in artificial intelligence, and on the contrary are essentially exploited the more precise biological neuron behaviors described in Sections Electrical synapse connections between neurons & Base model of an electrical synapse (gap junction) above.

One essential property to exploit is the intrinsic theoretical infinite parallelism offered by quantum mechanics[1], which allows to address the large number of upstream and postsynaptic links to a given synapse, and, beyond that, to mimic the networking of many neurons and synapses. The randomness of biochemical firings of the neuron links is mimicked by the randomness in qubit particles, of course at a different time scale, with experimental and theoretical justifications provided below.

The second essential property to exploit is the probabilistic nature of the quantum states which correspond to elements of randomness in synapse functionality and characteristics (Eqs. 2, 6).

The third property to exploit is the signaling and control features in synapses, which, using the signaling and control features in[12], might allow to introduce signaling and control in a quantum synapse gate model and a realization thereof (see Section Control and signaling in a quantum synapse).

We here assume that the quantum processing or sensing system incorporating quantum synapses has no intrinsic decay[36].

Some may question the validity of the "analogy" between the randomness of firings of the neuron links and the randomness in qubit particles, of course at different time scales. Their argument is that the neural firing randomness is driven by biochemical processes and stochastic variations, whereas quantum randomness results in superposition and measurement collapse. However, this argument ignores the randomness in the emissions at particle level of quantum particles, and also ignores that superposition and measurement / activation collapse happens as well in some biological processes at the level of specific biological functions like those mentionned in the Introduction. These processes indeed operate on vastly different principles, and thus cannot be treated as equivalent in signal processing terms at different time scales. However, with the experiments reported and analyzed in[13], and further specified by claims of Patent [A], when light or sound stimuli onto living species are made to follow a modulated Poisson distribution like that of a directly modulated semiconductor laser source useable as well as a quantum particle source, it was shown one can directly link in a probabilistic way the information bits carried by these stimuli to the measured titration of 17th ketosteroids in the animals, which represent a global indicator of neural activity in synapsis biochemical reactants (acetylcholine, adrenaline & noradrenaline). While very hard to measure, the neuron firing rate distributions are analyzed in[37] and linked to plasticity. So the said probabilistic correspondence is at information level, and this provides a partial justification to the chosen approach to be explored to design a "quantum synapse". This issue of "analogy" is further formalized in Section Limitations, results and open research issues.

In addition, for the further analysis of said biological processes, "quantum synapse" realizations may help analyze much faster the intricacies of said biological processes.

## Quantum synapse

The idea is to represent each of the electrical synapse links as a single component of a quantum state $|\psi>$, this applying respectively to both the upstream links 'k' $|\psi UP>$ to the presynaptic neurons, and to the downstream links 'l' $|\psi DOWN>$ to the postsynaptic neurons, thus the quantum system:

$$|\psi UP> = \sum_k a_k |\psi UP_k> \qquad (7)$$

$$|\psi DOWN> = \sum_l b_l |\psi DOWN_l> \qquad (8)$$

where:

- $\{|\psi UP_k>\}$, $\{|\psi DOWN_l>\}$ are two orthonormalized systems of eigenvectors;
- the values $|a_k|^2$, and $|b_l|^2$ are equal to the probabilities that the potentials VUP(t), VDOWN(t) of the presynaptic and postsynaptic neuron membranes, respectively, exceed the threshold potential $V_{thres}$. To simplify the readability, the corresponding quantum mechanics equations are not provided for these relations, as anyway the quantum measurements represent probabilities.

It is beyond the scope of this paper to make explicit the biological constraints on electrical potentials, and their differences, applicable to the synapse membranes, and originating in presynaptic and postsynaptic neuron potentials. It must only be pointed out that such bounds are stable over time.

The random sequence of presynaptic spikes (Eq. (2) (6)) is provided by the corresponding $|\psi UP>$ quantum state's randomness, and a logical CNOT gate function applied to the presynaptic neuron membrane potential down each link, the logical comparison being with $V_{thres}$.

The parallelism of links of the upstream (k) and postsynaptic (l) neuronal links is embedded in the dimensionalities of the vectors $|\psi UP>$, $|\psi DOWN>$.

We exploit both the two $|\psi UP>$ and $|\psi DOWN>$ quantum states' entanglement, and the quantum superposition of these states. Entangled states are those states of the quantum system, where qubits interact with each other, and involve a description of these states in a form wherein the product of wave functions of independent qubits is impossible.

These two state's evolutions are described by the time dependent equations:

$$|\psi UP> = CNOT (V(t) > V_{thres})) \qquad (9)$$

$$\text{CNOT qubit permutation matrix} = \begin{bmatrix} 1 & 0 & 0 & 0 \\ 0 & 1 & 0 & 0 \\ 0 & 0 & 0 & 1 \\ 0 & 0 & 1 & 0 \end{bmatrix}$$

$$i\hbar(\partial|\psi DOWN>/\partial t) = -(gL/Cm)\,(V(t)-V_{rest})\,|0> + i\hbar|\psi UP> \qquad (10\,a)$$

in which:

- $\hbar = h/2\pi$ is the Planck constant.
- $i = \sqrt{(-1)}$.

to which must be added the following constraints spelled out in Section Initial simple model :

$$(Constraint\ on\ total\ presynaptic\ neuron\ potentials) \qquad (10\,b)$$

$$(Constraint\ on\ total\ postsynapti\ neuron\ potentials) \qquad (10\,c)$$

The logical Controlled NOT (CNOT) functionality XOR's the first bit to the second bit and keeps the first bit unchanged. It has different optical or acoustic realizations[38], may be built from universal reversible gates[39], and may eventually be replaced by using a quantum Bell state[40], where a specific qubit may be measured as having a specific value with two probabilities adding up to one.

Note that here in (Eq. 10 a) we have $V_{rest}$, and in (Eq. 9) (and definition of the states) we have $V_{thres}$, due to the reinitialization of the membrane potential.

Figure 3. Single quantum synapse circuit, with one presynaptic and one postsynaptic neuron; the processes are those described in Eqs. 9–10 and Section Initial simple model; it is recalled that $V_{thres}$ is the threshold potential above which V(t) get's reinitialized to the value $V_{rest}$ approx. (60,5 mS/cm2, 50 mV).

## Control and signaling in a quantum synapse

In biology, synapses can be stimulated by brain and nervous commands through neurons, and end up affecting other nervous effectors or organs. The previous Sections pave the way to specify features whereby to modify the above quantum synapse model and circuit, so it renders effective a control functionality.

This rests on two properties, one of which is biological, and the other mathematical, applied in the context of quantum computing, and both with limitations:

- like biological synapses which are also subject to constraints, a quantum synapse can have quantum flows reproducing a two-way flow of signals, provided the quantum processing architecture itself is modified; this was mentioned in Section Electrical synapse connections between neurons and is exploited in Section Allowing for two-way junctions in some synapses;
- that general quantum processing signaling and control features as in[12] can capture control and selection functions by using some of the electrical synapse parameters of Section Quantum synapse as control variables; this is addressed in Section Implementation of signaling and control synapse qubits.

## Allowing for two-way junctions in some synapses

It was mentioned in Section Electrical synapse connections between neurons that, under some conditions, a biological electrical synapse can be transversed by bilateral flows of ions. Whereas the exploitation of two-way junctions is the theme of this Section, the biochemical description of the initial trigger effect by which ion flows become bidirectional in the synapse, belongs mostly to biochemistry and is outside the scope of this paper[37]. Below we also must make the assumption that the randomness of the downflows to the postsynaptic neuron is like the randomness of the upstream flows to the presynaptic neuron, so Eq. 6 can be reused.

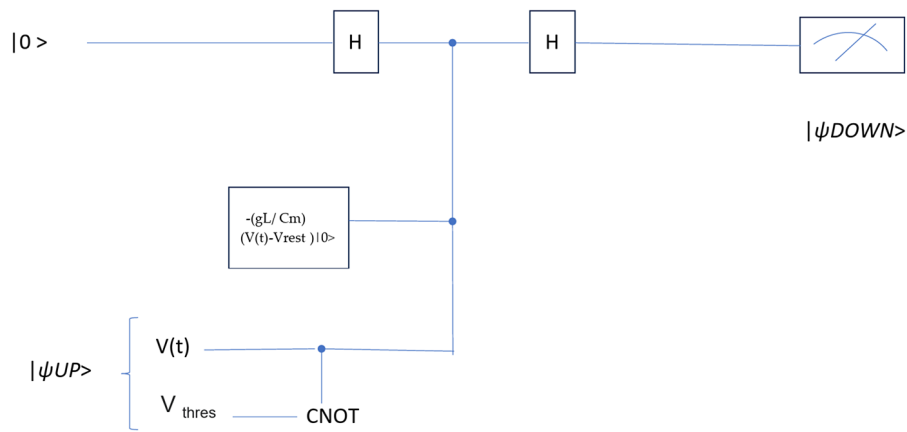

**Fig. 3**. Single quantum synapse circuit, with one presynaptic and one postsynaptic neuron; the processes are those described in Eq 9-10 and Section Initial simple model ; it is recalled that $V_{thres}$ is the threshold potential above which V(t) get's reinitialized to the value $V_{rest}$ approx. (60,5 mS/cm2, 50 mV).

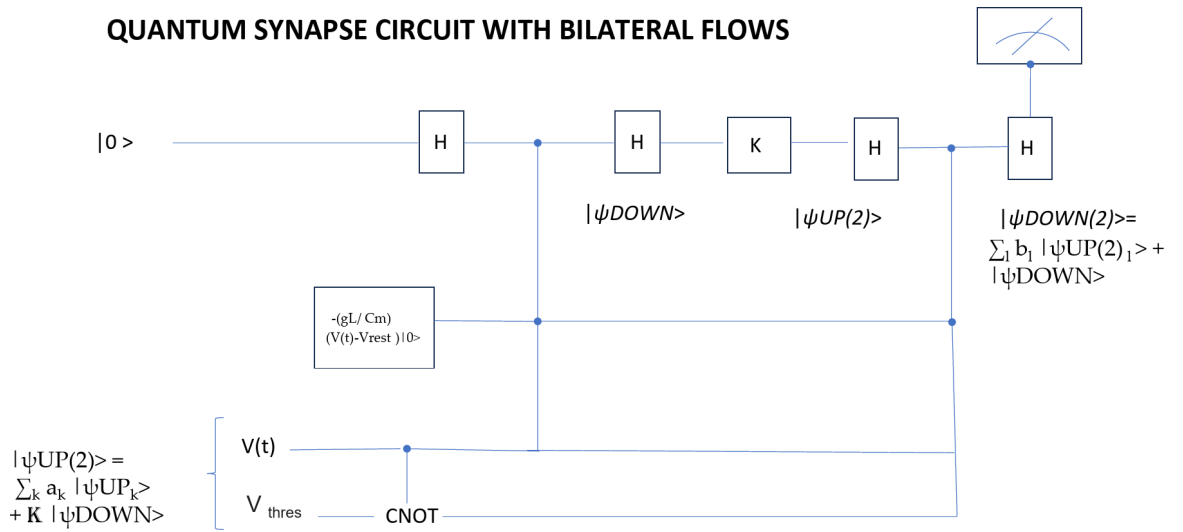

**Fig. 4**. Quantum synapse circuit with bilateral flows between one presynaptic neuron and one postsynaptic neuron.

The bidirectional functionality can be implemented by feeding a parametric part of the postsynaptic quantum state $|\psi DOWN>$, to mix with the input stream $|\psi UP>$ into the same synapse, the open Hermitian quantum operator **K** allowing for attenuation as well as phase shift; the resulting combined postsynaptic quantum state becomes $|\psi DOWN(2)>$ :

$$|\psi UP(2)> = \sum_k a_k |\psi UP_k> + \mathbf{K}|\psi DOWN> \tag{11}$$

$$ih\,(\partial|\psi UP(2)>/\partial t) = -(gL/Cm)\,(V(t) - Vrest) + ih|\psi DOWN> \tag{12}$$

$$|\psi DOWN(2)> = \sum_l b_l \left|\psi UP(2)_l\right> + \left|\psi DOWN\right> \tag{13}$$

where the $\{b_l\}$ characterize the postsynaptic electrical neuron membrane potentials, resulting from flows arriving from elsewhere to these neurons (see Section Quantum synapse).

The formalization and effects of the Hermitian quantum operator **K** are given in Appendix.

When determined, from measured biological constraints on the potentials mentioned above, the following corresponding additional constraints must be accounted for, and are here just mentioned for compliance:

$$(Constraint\ on\ total\ combined\ presynaptic\ neuron\ potentials) \tag{14}$$

$$(Constraint\ on\ total\ combined\ postsynaptic\ neuron\ potentials) \tag{15}$$

The result is a coupled set of equations, synthesized by the quantum gate circuit of Fig. 4. The quantum system operator **K** in effect represents in the simplest case a unitary matrix transform, alternatively a Hermitian matrix weighting, but extensions are possible to render it linked to probabilities of neuron potentials exceeding thresholds. Indeed, at the biophysical level, parts of potential in the postsynaptic neuron are fed back to add to the potential of the presynaptic neuron: attenuation, phase shift and possibly some delays may occur when they transverse the synapse. The so-called connexons, made up of sub-units called connexines (proteins), establish gap junction channels enabling the bidirectional flow of ions[41]. Research in biology has though not yet identified any cases with amplification; it has however established that electrical synapses with bidirectional flows enhance the speedy synchronization of potentials and electrical activity between groups of neurons.

This shows, with reference to Section Quantum synapse, that the qubit attributes with complex values $\{a_k\}$, and $\{b_l\}$ can, respectively, be considered to be control parameters to the electrical synapse from presynaptic neurons, and considered to be the resulting control parameters to postsynaptic neurons. If their absolute values are equal to zero, the corresponding links to the corresponding neurons are shut down. This property is the key justification for an embedded quantum synapse used in quantum sensing and control.

### Implementation of signaling and control synapse qubits

It was shown in[12,13], both in mathematical terms but also in terms of realization by manipulating magnetism by electrical fields, that the quantum pathways can be assigned to be linked either to information carrying qubits, or to quantum signaling or control gates which have eigenvector attribute tags like $\{a_k\}$ or $\{b_l\}$ assigned by a predecessor quantum gate providing said signaling or control. If furthermore these tags possess the property of being elements of a colored algebra, decidable quantum signaling can be achieved. This is essential in some applications. The only drawback is to extend the quantum state space, and, for those quantum gates serving for control or signaling, to have additional specific quantum gate realizations of those.

This leads to the following idea, for a quantum synapse, to have:

i. a unique mapping of the chemical calcium signaling ion concentration levels, described above and in[42,43], into unique elements of a colored algebra, so that even binding rules of two Ca2 + ions can map into algebraic rules for algebraic color change, the chemical details of which are beyond the scope of this paper[44];
ii. decompose some of the vector dimensions of the qubits |ψUP(2) > and |ψDOWN(2)>, to have some of the corresponding attributes (like $\{a_k\}$'s or $\{b_l\}$'s) carry said decidable tags.

However, it is not within the scope of the present paper to describe in detail the biochemical properties and the mix of the ion flows flowing through an electrical synapse.

## Quantum sensing evaluation application

One of several application fields, as envisaged for evaluation, is briefly discussed. The author has worked for a long time on mission critical sensor fusion architectures and specific sensors[45–47], to realize over time the bottlenecks rooted in geometrical information matching, in real time sensor tuning and adaptation, and in probabilistic information aggregation, even when using high performance computing. Quantum computing mimicking biological functions offers vastly superior potential due to computational speed, state superposition, entanglement, hypothesis refinement and probabilistic assessments, as the split/merge/evaluate functionalities can be coordinated by a quantum synapse over a high number of quantum flows, largely better than a blackboard with its read/write/store and synchronization problems. Once the first quantum synapse circuits have been realized in hardware, a sensor fusion experiment will be conducted, probably on a quantum processing element itself used for sensor operations optimization (magnetic fields, electrical fields, thermal characteristics, etc.…). Extensive simulations have already been done though.

## Discussion on scalability

The proposed quantum synapse models of Sections Quantum synapse & Control and signaling in a quantum synapse, resting on the base model of Section Base model of an electrical synapse (gap junction) , involves some quantum gates, each gate processing the quantum state having to operate sequentially if traditional computing was to be used. Under this last hypothesis, serious scalability concerns could be raised, as each synapse would require significant time to compute its quantum states. In a biological network of neurons each synapse has around 10,000 postsynaptic neurons, meaning the computational cost to simulate an entire neuromorphic system might then be significant. The time required to calculate quantum states and their probabilities for a large number of synapses would quickly become prohibitive.

However, this paper explores an alternative approach: instead of calculating sequentially all intermediate quantum states, quantum processing with its parallelism allows to mimic as explained above the spike trains as random quantum particles, and exploits parallelism, and thus to estimate by quantum measurements the new synapse states with probabilities as explained in Section Quantum synapse and Allowing for two-way junctions in some synapses. This largely simplifies the computational requirements and makes the model more feasible for large-scale applications when using quantum computing.

## Quantum synapse simulation

Quantum computing simulations of the two designs (Figs. 3, 4) have been conducted and are under analysis in view of later optimization for execution on a quantum computer as well as first on an array of eight multicore FPGA's. The main simulation tool has been MIMIQ[48]based on research at the European center for quantum sciences, relying on Matrix product states. It enables linked partial differential equations as well as combinatorics. Students associated with the project in advance get training on the QLM simulator[49,50]with initial code specified in Python with cross-assemblers. Of special interest, and also quite complex, is the first understanding and possible exploitation of the superposition and imbrication between the presynaptic neurons and the postsynaptic neurons. Next, critical is the exploitation of the state of the art for noise and randomness in biological synapse operations (Eq. (2), Eq. (5))[37], and their comparison with the noise and errors in the quantum computations on a quantum computer; thus much efforts have gone into the design of parametrizations of random number generators (parallelled with quantum particle source generators). Finally, at this stage of the research, has been simulated in a limited way the proliferation of presynaptic and postsynaptic neurons for a single quantum synapse. At all simulation stages, the time-dependent simulation is let to lapse with varying number of simulation cycles as required by quantum computations, to see how the quantum synapse itself evolves over time (e.g. for the Example of Section Quantum sensing evaluation application), although the time scales are obviously completely different than for biological synapses ; this requires very extensive memory capabilities (32 Tb). Parameters and results will be provided in a companion paper.

### Limitations, results and open research issues

The claim is not for a probabilistic equivalence between synapse firing stochastics, and quantum particle stochastics; both have stochastic distributions which can be characterized for a given neural condition and a given quantum physical set-up & particle source. Therefore, should be characterized the approximate matching between these two classes of stochastic distributions from calibration data with:

- for a synapse, the stochastic distribution of the neuron membrane potential along a given synapse link V(t), as modelled by Eq. (1);
- for a quantum synapse design Figs. 3 or 4), the stochastic distributions of the values $|a_k|^2$, and $|b_l|^2$ of the probabilities that the potentials VUP(2)(t), VDOWN(2)(t) of the presynaptic and postsynaptic neuron membranes, respectively, exceed the threshold potential $V_{thres}$, after traversing a quantum circuit realization .

One limitation in the present "quantum synapse" design is that it does not capture all the details and versatility of true synapses, realizing that this is not the research goal of this paper (Section Research question ) and that biology research on such questions is very detailed. Likewise, synapses are not the only cells involved in the interactions between neurons and nerves. It is e.g., to be recalled that glia cells (astrocytes and oligodendrocytes[51]) play a role in neuro-muscular interfaces, and that they contribute to the spatial distribution of connectivity between neurons. Such aspects are beyond the scope of this article.

Another limitation is that this article cannot include initial experimental results beyond those stated in[13], and quantum simulation results from Section Quantum synapse simulation which will be reported in a follow-up publication. The time scale of the stochastic process of Eq. 2 has obviously been adjusted to fit quantum particle flows with on-going quantum particle source selection, and the biological parameters reported in Section Base model of an electrical synapse (gap junction) were adjusted accordingly. It was observed that for some structures of the open quantum operator **K** (Eq. 11) remarkably interesting spatiotemporal filtering was obtained.

A remaining research question is to speed up the particle physics realizations and measurements according to the plan already outlined in[5], and to narrow down the functionalities to be built to those outlined here, especially the quantum paths and their control in a quantum synapse.

### Conclusion

This paper has attempted to pave the way towards an initial quantum circuit design for an electrical synapse, including its dynamics, with a progression from a simple biological model to a quantum processing model and quantum gate designs, with parametric bi-directional flows and signaling.

The discussion around a quantum sensing application to sensor fusion for quantum processing element monitoring, demonstrates the capacity of a quantum synapse gate to integrate sensors and quantum sensors. This quantum synapse can perform bidirectional flows and leverage quantum computing principles such as superposition and entanglement to enhance sensing capabilities.

Furthermore, the proposed quantum synapse circuits may fit embedded systems architectures with both control, sensing, and probabilistic computations. This theoretical potential may eventually over time be slowly scaled against the capabilities of a human brain of 1400 g, with approx. 10 000 synapses per neuron and 85–100 Billion neurons, whereby each presynaptic neuron potentially interacts with 10 000 postsynaptic neurons[52], so the scaling up of quantum synapses has a long way to go. Consequently, a more interesting goal may be to analyze and achieve adaptation, thus the need for signaling and control.

It should be highlighted that the chosen approach with LIF synapses contrasts with the training of neural networks on quantum computers, which is essentially a recursive regression on very simple neuron approximations without synapse functionalities.

### Data availability

This manuscript does not report data generation or analysis other than those reported or from listed references. The datasets used and/or analysed during the current study are available from the corresponding author on reasonable request. Data and Results for the cited quantum simulation of a quantum synapse will appear in a different article.

### Appendix

The quantum operator **K** from Eq. (11) is in general Hermitian. This means by definition, that this operator **K** in a Hermitian space H, obeys the property: $U$ (x, y) ε $H^2$, (**K**(x)|y) = (x |**K**(y)). Hermitian operators play an important role in quantum mechanics because they represent physical quantities such as attenuation, phase shift, delays. The real eigenvalues represent possible values of the relative strengths of variables in |ψDOWN>, and the eigenfunctions (or vectors) represent the associated states thereof. Hermitian operators allow to describe observables, like voltages and energy (see Section Quantum synapse). They ensure that the measurement values are real numbers, which is essential for the coherence of physical predictions in the presence of more than a pair of synaptic links. A set of Hermitian operators which commute amongst them, each associated with several synaptic links, represent a complete set of observables, allowing for the simultaneous measurement of several physical quantities, here several synaptic link properties. A particular class of Hermitian operators **K** are quantum rotation operators R, which allow us here to represent phase differences between several synaptic links. With every physical rotation R, a quantum mechanical rotation operator Θ ( R): H -> H, is the rule that assigns to each vector in the space H the state vector $|α>_R$ = Θ(R)|α > that is also in H. Furthermore Θ (**Ω**, φ) = exp (-iφ**Ω**. **J**/h) wherein **Ω** is the rotation axis, φ the rotation angle, **J** is the angular momentum operator, and h is the reduced Planck constant. Such quantum mechanical rotation operators help in the quantum disambiguation: contrary to classical rotations in a Euclidian space, quantum rotations are non-commutative, which means that the order by which the phase shifts are measured affects the final state of the system. At the same time, the quantum rotation operator maintains rotational symmetry, which is essential to quantify the quantum angular moment, and it allows to manipulate the spin of an electron. For further theoretical justifications, see e.g.: L.D. Landau and E.M. Lifshitz: *Quantum Mechanics: Non-Relativistic Theory*, New York: Pergamon Press, 1985.

### Acknowledgements
The author thanks for the prolific discussions and debates with several advanced research programs, laboratories, or visionaries in industry worldwide about simulations, component realizations and operations. This includes Prof. P.N. Borza, Transylvania University, with whom parts of the underlying biology inspired quantum processing architecture has been developed. Patent [A] and PCT patent applications are pending.

### Author contributions
LFP wrote the whole submission.

### Declarations

### Competing interests
The authors declare no competing interests.

### Patents
A. Restian, A.; Borza, P.N.; Daghie, M.V.; Nicolau, N. Metoda si aparat de investigare a influentei solicitarilor informationale esupra organismului (Method and apparatus for investigating the influence of information on living organisms). Romanian Patent Invention claim no 119 168 of 17 June 1985, assigned to Institutul de igiena si sanatate publica, Bucarest. Romanian Patent no 93122 (24 June 1987)

### Additional information
**Correspondence** and requests for materials should be addressed to L.-F.P.

**Reprints and permissions information** is available at www.nature.com/reprints.

**Publisher's note**  Springer Nature remains neutral with regard to jurisdictional claims in published maps and institutional affiliations.